\documentclass[11pt]{article}

\makeatletter

\@addtoreset{equation}{section}

\makeatother

\newcommand{\calB}{{\cal B}}

\newcommand{\calF}{{\cal F}}
\newcommand{\calG}{{\cal G}}
\newcommand{\calH}{{\cal H}}
\newcommand{\calI}{{\cal I}}

\newcommand{\calO}{{\cal O}}
\newcommand{\calP}{{\cal P}}

\newcommand{\bC}{{\bf C}}

\newcommand{\bR}{{\bf R}}
\newcommand{\bZ}{{\bf Z}}

\newcommand{\trace}{{\rm tr}}
\newcommand{\diam}{{\rm diam}}

\newcommand{\la}{\langle}
\newcommand{\ra}{\rangle}
\newcommand{\cube}{\Delta}
\newcommand{\bi}{\hspace{-1cm}}

\newtheorem{theorem}{Theorem}[section]
\newtheorem{prop}[theorem]{Proposition}
\newtheorem{lemma}[theorem]{Lemma}
\newtheorem{corollary}[theorem]{Corollary}
\newtheorem{remark}[theorem]{Remark}
\newtheorem{example}[theorem]{Example}

\newcommand{\proof}{\noindent {\em Proof:~}}  

\newcommand{\nn}{\nonumber}

\def\bd{\begin{displaymath}}
\def\ed{\end{displaymath}}

\def\eqref#1{(\ref{#1})} 
\def\qed{\hbox{\hskip 6pt\vrule width6pt height7pt depth1pt
    \hskip1pt}\bigskip}  

\def\to{\rightarrow}
\def\ltoi{{\Lambda \nearrow \bZ^d}}

\def\runinend{\enspace}
\def\ackname{Acknowledgment\runinend}%
\def\acknowledgments{\par\addvspace{17pt}\rmfamily
\def\ackname{Acknowledgments\runinend}%
\trivlist\if!\ackname!\item[]\else
\item[\hskip\labelsep
{\bf\ackname}]\fi}%

\begin{document} 
\bibliographystyle{plain} 

\hfill{} 
 
\thispagestyle{empty}

\begin{center}{\bf \Large Large Deviations in Quantum Lattice Systems:\\
\smallskip
One-phase Region } 
\vspace{6mm}

\large{\bf Marco Lenci} 

{\small\it Department of Mathematical Sciences, Stevens Institute of
Technology, \\ Hoboken, NJ 07030, Email: mlenci@math.stevens.edu \\ }

\vspace{5mm}

\large{\bf Luc Rey-Bellet}

{\small\it Department of Mathematics and Statistics, University of 
Massachusetts,\\ Amherst, MA 01003, Email: lr7q@math.umass.edu \\ }

\end{center}

\vspace{-1mm}

\setcounter{page}{1} 
\begin{abstract}  
We give large deviation upper bounds, and discuss lower bounds, for
the Gibbs-KMS state of a system of quantum spins or an interacting
Fermi gas on the lattice. We cover general interactions and general
observables, both in the high temperature regime and in dimension one.
\end{abstract}

\section{Introduction and statement of the results} 

The study of large deviations of macroscopic observables plays a
fundamental role in classical statistical physics, for example in the
study of the equivalence of ensembles and in hydrodynamical limits.
The large deviation principle for Gibbs measures in classical
mechanics, initiated in the seminal papers \cite{Ru65,La} is now a
classical subject, see e.g.\ \cite{El, Ol, Co, FoOr, Ge, DeStZe,
LePfSu1, LePfSu2, Pf}.  It is maybe surprising that, in comparison,
very little is known about large deviations in quantum statistical
mechanics. To our knowledge, results on large deviations have been
obtained only for the fluctuations of the particle density in
\cite{LeLeSp}, for ideal fermionic and bosonic quantum gases, and in
\cite{GaLeMa}, for dilute fermionic and bosonic gases in (using
cluster expansion techniques). As we were completing this paper a
preprint \cite{NeRe} appeared where large deviation results for
observables that depend only on one site are established in the high
temperature regime (using again cluster expansions). In this work,
however, we consider general observables. Also, from a technical point
of view, we do not use cluster expansions, but a simple matrix
inequality, combined with analyticity estimates on the dynamics and
subadditivity arguments.

Other large deviation results in quantum mechanical models can be
found in \cite{BeLePu1, BeLePu2, OhPe, RaWe1, RaWe2}. In the larger
context of probabilistic results for quantum lattice systems we want
to mention \cite{GoVeVe1, GoVeVe2, Ma1, Ma2}, about the Central Limit
Theorem and the algebra of normal fluctuations, and \cite{BKSS1,
BKSS2}, on Shannon--McMillan type of theorems.

For simplicity we will consider quantum systems of spins or fermions
on a lattice.  The statistical mechanics of spins and fermions is
naturally expressed in the formalism of $C^*$-algebras, which we will
use throughout the paper (refer to \cite{BrRo, Si} for spin systems
and to the recent \cite{ArMo} for fermions).  For classical Gibbs
systems, the DLR condition is a crucial ingredient in establishing the
large deviations principle. For quantum systems, Araki introduced an
analogous condition, the so-called Gibbs condition, which will play an
important role in our proofs. Our results presumably extend to some
bosonic systems or lattices of oscillators, but such systems present
more technical difficulties as they are most naturally expressed in
the $W^*$-algebraic formalism.
 
In order to illustrate the scope---and limitations---of this work, let
us consider an example of a system of (fermionic) particles of spin
$1/2$. Our results, however, are more general and will be detailed in
Section \ref{qmgf}.

Let $c_{x,\sigma}$ and $c^*_{x,\sigma}$ denote the annihilation and
creation operators for fermions of spin $\sigma \in
\{\uparrow, \downarrow\}$ at site $x \in \bZ^d$. These operators
satisfy the Canonical Anticommutation Relations. We denote by
$n_{x,\sigma} = c^*_{x,\sigma} c_{x,\sigma}$ the operator for the
number of particles in $x$ with spin $\sigma$, so that $n_x =
n_{x,\uparrow} + n_{x,\downarrow}$ indicates the operator for the
total number of particles in $x$. The finite-volume Hamiltonian (with
free boundary conditions) for a finite subset $\Lambda$ of the lattice
is taken to be
\begin{eqnarray} \label{hamex}
H_\Lambda &=& - \sum_{\{x,y\} \subset \Lambda} T_{x-y} \sum_\sigma
\left( c^*_{y,\sigma} c_{x,\sigma} + c^*_{x,\sigma} c_{y,\sigma}
\right) + \sum_{x \in \Lambda} U n_{x,\uparrow} n_{x,\downarrow} \nn\\
&& + \sum_{\{x,y\} \subset \Lambda} J_{x-y} \, n_x n_y \,.
\end{eqnarray}
Special cases of this Hamiltonian are the Hubbard models and the
$tJ$-models, which are widely used in applications. We define the
number operator for a finite subset $\Lambda$ of the lattice by
\begin{equation}
N_\Lambda \,=\,  \sum_{x\in \Lambda} n_x \,.
\end{equation}
We work in the grand canonical ensemble, whose finite-volume Gibbs
states are given by
\begin{equation}
\omega^{(\beta, \mu)}_\Lambda( \: \cdot \:) \,=\, \frac{\trace ( \:
\cdot \:\, e^{-\beta(H_\Lambda-\mu N_\Lambda)} )} {\trace (
e^{-\beta(H_\Lambda-\mu N_\Lambda)} )} \, ,
\end{equation}
where $\beta$ is the inverse temperature and is the chemical potential
$\mu$. We denote by $\omega^{(\beta,\mu)}$ the Gibbs states of the
infinite systems, which, roughly speaking, are constructed as limit
points of $\omega_\Lambda$ as $\Lambda \nearrow \bZ^d$. Mathematically
the Gibbs states of the infinite system are characterized,
equivalently, by either the variational principle, the KMS condition,
or the Gibbs condition. 

Let us consider a local (microscopic) observable $A$, i.e., a
self-adjoint operator depending only on even products of creation and
annihilation operators (this is a natural limitation for Fermi
systems; see Section \ref{qlatt}). For example, say that $A$ is an
even polynomial of the creation and annihilation operators.  Assume
that $A$ depends only on the sites of $X \subset \bZ^d$, and denote by
$\upsilon_x A$ the translate of the operator $A$ by $x \in \bZ^d$.
For a cube $\Lambda$ we define the macroscopic observable $K_\Lambda$
by
\begin{equation}
K_\Lambda\,=\, \sum_{X+x \subset \Lambda} \upsilon_x A \, ,
\end{equation}
In this way, $|\Lambda|^{-1} K_\Lambda$ represents the average of $A$ in
$\Lambda$ ($|\Lambda|$ denotes the cardinality of the set $\Lambda$).  
If we denote by ${\bf 1}_B$ the indicator function of a Borel
set $B \subset \bR$, then, for a Gibbs state $\omega^{(\beta, \mu)}$,
\begin{equation}\label{rolam}
\rho_{\Lambda} (B) \,= \, \omega^{(\beta,\mu)} 
\left( {\bf 1}_B \left(|\Lambda|^{-1} K_\Lambda\right)\right)
\end{equation}
defines a probability measure on $\bR$. Namely, (\ref{rolam}) is the
probability that, in the state $\omega^{(\beta,\mu)}$, the observable
$|\Lambda|^{-1} K_\Lambda$ takes values in $B$. Large deviation theory
studies the asymptotic behavior of the family of measures
$\rho_{\Lambda}$ on an exponential scale in $|\Lambda|$.  This
asymptotic behavior is expressed in terms of a rate function $I(x)$,
which is a lower continuous function with compact level sets.

Let $C$ be a closed set. We say that we have a large deviation upper
bound for $C$ if 
\begin{equation}\label{ldup}
\limsup_{ \ltoi} \frac{1}{|\Lambda|} \log \omega^{(\beta,\mu)} \left(
{\bf 1}_C \left(|\Lambda|^{-1} K_\Lambda\right)\right) \le - \inf_{x
\in C} I(x) \,.
\end{equation}
Similarly, if $O$ is open, we have a large deviation lower
bound for $O$ if 
\begin{equation}\label{ldlo}
\liminf_{\ltoi} \frac{1}{|\Lambda|} \log \omega^{(\beta,\mu)} \left(
{\bf 1}_O \left(|\Lambda|^{-1} K_\Lambda\right)\right) \ge - \inf_{x
\in O} I(x)\,.
\end{equation}
One says that $\{ \rho_\Lambda \}$ satisfies the large deviation
principle if we have upper and lower bound for all closed and open
sets, respectively.

In order to study the large deviations for $\rho_{\Lambda}$ as $\ltoi$
(along sequences of cubes), one considers the corresponding logarithmic
moment generating function, defined as
\begin{equation}\label{fal}
f(\alpha) \,=\, \lim_{\ltoi}  \frac{1}{|\Lambda|} \log 
\omega^{(\beta,\mu)}( e^{\alpha K_\Lambda})  \,.
\end{equation}
The G\"artner--Ellis Theorem (see, e.g., \cite{DeZe}), shows that the
existence of $f(\alpha)$ implies large deviation upper bounds with a
rate function $I(x)$ that is the Legendre transform of $f(\alpha)$.
One obtains lower bounds if, in addition, the moment generating
function is smooth, at least a $C^1$. If the moment generating
function is not smooth, one has a weaker result: in \eqref{ldlo}, the
infimum over $O$ is replaced by the infimum over $O \cap E$, where $E$
is the set of the so-called exposed points (see \cite{DeZe} for
details).

Our results apply both in one dimension and at high temperature. In
both cases the parameters $\beta$ and $\mu$ are such that there is a
unique Gibbs-KMS state $\omega^{(\beta,\mu)}$.

\medskip
\noindent 
{\bf Dimension one.}  Let us assume that the lattice is
one-dimensional and that the interaction has finite range, i.e., there
exists an $R>0$ such that $T_{x-y} $ and $J_{x-y}$ vanish whenever
$|x-y|>R$.  Our core result is that, for any macroscopic observable
$K_\Lambda$ and all values of $\beta$ and $\mu$, the moment generating
function $f(\alpha)$ exists and is finite for all $\alpha \in \bR$.
Furthermore $f(\alpha)$ is given by the formula
\begin{equation} \label{form} 
f(\alpha) \,=\, \lim_\ltoi \frac{1}{|\Lambda|} \log \frac{ \trace (
e^{\alpha K_\Lambda} \, e^{-\beta (H_\Lambda-\mu N_\Lambda)} ) }
{\trace ( e^{-\beta(H_\Lambda-\mu N_\Lambda)} )} \,,
\end{equation}
which involves only finite-dimensional objects.

As recalled above, if $I(x)$ is the Legendre transform of $f(\alpha)$,
the G\"artner--Ellis Theorem entails the large deviation upper bounds
with $I$ as the rate function.  As for the lower bounds, it is
tempting to conjecture that the function $f(\alpha)$ is smooth, in one
dimension. We have not proved it so far.

It is instructive, at this point, to compare our results on quantum
systems with their classical analogs.  In the classical case, using
the DLR equations, one shows that a formula similar to
Eq.~\eqref{form} holds, with the trace replaced by the expectation
with respect to the counting measure.  In that case, one sees that
$f(\alpha)$ is simply the translated pressure corresponding to the
Hamiltonians $\beta H_\Lambda - \beta \mu N_\Lambda - \alpha
K_\Lambda$. Therefore, classically, the smoothness of $f(\alpha)$
follows immediately from the lack of phase transitions in one
dimension, together with the identification of Gibbs states with
functionals tangent to the pressure \cite{Is, Si}.  In the quantum
case, $K_\Lambda$ does not commute with $H_\Lambda-\mu N_\lambda$, in
general, so the thermodynamic interpretation of the moment generating
function is not obvious. This is the main difference.

Such interpretation is possible, however, when $K_\Lambda$ commutes
with $H_\Lambda - \mu N_\Lambda$.  In our example (\ref{hamex}),
$H_\Lambda$ does commute with $N_\Lambda$ (this is not hard to verify,
using the CAR; see (\ref{car})), so that we can fully treat the
physically important large deviations in the energy and in the
density.  More in detail, if the pressure function for our system is
defined as
\begin{equation}
P(\beta,\mu) \,=\, \lim_\ltoi \frac1{|\Lambda|} \log \trace
\left(e^{-\beta(H_\Lambda - \mu N_\Lambda)} \right) \,,
\end{equation}
then (\ref{ldup})-(\ref{ldlo}) hold for the energy ($K_\Lambda =
H_\Lambda$), with $I(x)$ being the Legendre transform of
\begin{equation} \label{trpren}
\alpha \,\mapsto\, P \left( \beta-\alpha, \frac{\beta}{\beta -\alpha}
\mu \right) - P(\beta, \mu)\,.
\end{equation} 
They also hold for the number of particles ($K_\Lambda =
N_\Lambda$), and in that case $I(x)$ is the Legendre transform of
\begin{equation} \label{trprde}
\alpha \,\mapsto\, P \left( \beta, \mu + \frac{\alpha}{\beta} \right)
- P(\beta, \mu)\,.
\end{equation}

\medskip
\noindent 
{\bf High temperature.}  For arbitrary space dimension we assume that
that the interaction is summable: $\sum_{x\in \bZ^d} |T_{x}| +|J_{x}|
<\infty$. Our main result is that there exist two constants, $\beta_0$
(which depends only on the Hamiltonians $H_\Lambda$) and $\alpha_0$
(which depends only on the observable $K_\Lambda$), such that the
function $f(\alpha)$ exists for $|\alpha| < \alpha_0$, $|\beta| <
\beta_0$, and arbitrary $\mu \in \bR$.  Furthermore, in the special
case in which the macroscopic observable is a sum of terms depending
only on one site, $\alpha_0$ can be taken to be infinity. Again, the
function $f(\alpha)$ is also given by \eqref{form}.

This yields large deviation upper bounds for closed sets which are
contained in a neighborhood of the average $\overline{K} = \lim_\ltoi
|\Lambda|^{-1} \omega^{(\beta,\mu)}(K_\Lambda)$.  At high temperature,
one expects $f(\alpha)$ to be smooth, in fact analytic, and this can
be proved using a cluster expansion \cite{LeRe}.

For the case of commuting observables we show that the moment
generating function exists for any $\alpha$, provided $|\beta| <
\beta_0$.  It is known that, for sufficiently high temperature and any
value of the chemical potential, there is a unique Gibbs state (see
Theorem 6.2.46 of \cite{BrRo}).  Using this, we obtain a full large
deviation principle for the particle number (or density). As for the
energy, we expect $f(\alpha)$ to have a singularity at some $\alpha
\ne 0$; at any rate, we have upper bounds for all closed sets and
lower bounds for sets that are contained in a neighborhood of the mean
energy.  For both the energy and the particle density, the rate
functions are again the Legendre transforms of
(\ref{trpren})-(\ref{trprde}).

\vspace{0.2cm}
\noindent
Once again, the precise statements for a general quantum lattice
system will be presented---and proved---in Section \ref{qmgf}.

\section{Quantum lattice systems} \label{qlatt} 

We consider a quantum mechanical system on the $d$-dimensional lattice
$\bZ^d$, as seen, e.g., in \cite{Ru, Is, BrRo, Si} for spin systems,
and in \cite{ArMo} for fermions.

\subsection{Observable algebras}

We first describe quantum spin systems. Let $\calH$ be a
finite-dimensional Hilbert space. One associates with each lattice
site $x \in \bZ^d$ a Hilbert space $\calH_x$ isomorphic to $\calH$ and
with each finite subset $X \subset \bZ^d$ the tensor product space
$\calH_X \,=\, \bigotimes_{x\in X} \calH_x$.  The local algebra of
observables is given by $\calO_X= \calB(\calH_X)$, the set of all
bounded operators on $\calH_X$. If $X \subset Y$, there is a natural
inclusion of $\calO_X$ into $\calO_Y$, and the algebras $\{ \calO_X
\}$ form a partially ordered family of matrix algebras. The
norm-completion of the union of the local algebras is a $C^*$-algebra
denoted by $\calO$ which correspond to the physical observables of the
system. In particular we have that $[\calO_X\,,\, \calO_Y] =0$
whenever $X \cap Y = \emptyset$.  A state $\omega$ is a positive
normalized linear functional on $\calO$, i.e., $\omega\,:\, \calO
\longrightarrow \bC$, $\omega({\bf 1}) = 1$ and $\omega(A) \ge 0$,
whenever $A \ge 0$.  The group $\bZ^d$ acts as a $^*$-automorphic
group on $\calO$: For $x\in \bZ^d$, $\upsilon_x (\calO_X) \,=\,
\calO_{X+x}$.  A state is called translation invariant if $\omega
\circ \upsilon_x = \omega$ for all $x \in \bZ^{d}$ and we denote by
$\Omega_{I}$ the set of all translation invariant states.  The action
of $\upsilon$ is asymptotically abelian: therefore $\Omega_{I}$ is a
Choquet simplex and one can decompose a state into ergodic components
(see \cite{Si}).

The structure of the algebra of observables for fermionic lattices
gases is a little more involved, due to the anticommutativity
properties of creation and annihilation operators (see
\cite{ArMo,DaFeFrRe}). We construct it as follows.

Let $\calI$ be the finite set that is supposed to describe the spin
states of a particle. For $X$ a finite subset of $\bZ^d$, $\calF_X$ is
defined formally as the $C^*$-algebra generated by the elements $\{
c^*_{x,\sigma},\, c_{x,\sigma} \}_{x\in X, \sigma\in \calI}$ together
with the relations
\begin{eqnarray} 
\{c^*_{x,\sigma} \,,\, c_{y,\sigma'} \} \,&=&\, \delta_{x,y} 
\delta_{\sigma,\sigma'} 
{\bf 1} \nn \\
\label{car}
\{c^*_{x,\sigma} \,,\, c^*_{y,\sigma'} \} \,&=&\,
\{c_{x,\sigma} \,,\, c_{y,\sigma'} \} \,=\, 0 \,.
\end{eqnarray} 
The above are referred to as CAR (Canonical Anticommutation
Relations).  $c^*_{x,\sigma}$ and $c_{x,\sigma}$ are called the
annihilation and creation operators and are taken to be mutually
adjoint by definition. It is easy to realize that, as a vector
space,
\begin{equation}
\calF_X \,=\, {\rm span} \left\{ c^{\sharp_1}_{x_1, \sigma_1} \,
c^{\sharp_2}_{x_2, \sigma_2} \cdots \, c^{\sharp_m}_{x_m, \sigma_m}
\right\},
\label{calF_X}
\end{equation}
where the span is taken over all (finite) sequences $\{
(x_j,\sigma_j,\sharp_j) \}_{j=1}^m$ in $X \times \calI \times \{
\cdot, * \}$ that are strictly increasing w.r.t.\ a predetermined
order. If $X \subset Y$, there is a natural inclusion $\calF_X \subset
\calF_Y$, and we define the fermionic $C^*$-algebra $\calF$ to be the
norm-completion of $\bigcup_{X \subset \bZ^d} \, \calF_X$.

Elements of $\calF$ localized on disjoint parts of the lattice do not
necessarily commute (they might either commute or anticommute) and so
$\calF$ is not asymptotically abelian.  We have to restrict the class
of allowed observables to a smaller algebra. Let us denote by $\Theta$
the automorphism of $\calF$ determined by $\Theta( c^\sharp_{x,\sigma}
) = - c^\sharp_{x,\sigma}$.  The observable algebra of a fermionic
lattice gas $\calO$ is defined to be the even part of $\calF$, i.e.,
\begin{equation}
\calO \,=\, \left\{ A \in \calF \,|\, \Theta(A) = A \right\} .  
\end{equation} 
Clearly, $\calO_X = \calO \cap \calF_X$ is given by the same r.h.s.\
of (\ref{calF_X}), restricted to $m$ even. Hence $[\calO_X\,,\,
\calF_Y] = 0$ whenever $X \cap Y = \emptyset$, which is the
commutativity property we need. The algebra $\calO$ is thus quasilocal
and similar considerations as for quantum spins systems apply.

\noindent
\begin{example}\label{s12}
{\rm   
For quantum spin systems with spin $1/2$, the Hilbert spaces 
$\calH_x$, $x\in \bZ^d$ is isomorphic to $\bC^{2}$}. 
\end{example}

\begin{example}\label{f12}{\rm 
For fermionic systems of particles with spin $1/2$, for each $x$, the
algebra generated by $c^*_{x,\sigma}$ and $c_{x,\sigma}$ is isomorphic
to $\calB(\bC^4)$.  }
\end{example}

\subsection{Interactions and macroscopic observables} \label{interactions}

An interaction $\Phi = \{ \phi_X \}$ is a map from the finite subsets
$X$ of $\bZ^d$ (denoted $\calP_f (\bZ^d)$) into the self-adjoint
elements of the observable algebras ${\calO}_X$ (denoted
$\calO_X^{(sa)}$).  We will always assume the interaction to be
translation invariant, i.e., $\upsilon_x \phi_X = \phi_{X+x}$ for all
$x \in \bZ^d$ and all $X \in \calP_f (\bZ^d)$. An interaction is said
to have finite range if there exists an $R>0$ such that $\phi_X=0$
whenever $\diam(X)$, the diameter of $X$, exceeds $R$. (One usually
says that the range is $R$ if $R$ is the smallest positive number that
verifies the previous condition.)  We denote by $\calB^{(f)}$ the set
of all finite range interactions. The set of interactions can be made
into a Banach space by completing $\calB^{(f)}$ with respect to
various norms.  In this paper we use the norm
\begin{equation}\label{thenorm}
\|\Phi\|_\lambda \,=\, \sum_{X \ni 0} \|\phi_X\| \, e^{\lambda |X|}
\,,
\end{equation}
where $\lambda > 0$ and $|X|$ denotes the cardinality of $X$. We
call $\calB_\lambda$ the corresponding Banach space of
interactions. To a given $\Phi$ one associates a family of
Hamiltonians (or energy operators) $\{ H_\Lambda \}_{\Lambda \in
\calP_f (\bZ^d)}$ via
\begin{equation}
H_\Lambda = H_\Lambda (\Phi) = \sum_{X \subset \Lambda} \phi_X.
\label{H_L}
\end{equation}

As in \cite{La}, we define a finite-range macroscopic observable $K$
of range $R$ to be a mapping $K \,:\,\calP_f (\bZ^d) \longrightarrow
\calO^{(sa)}$ such that
\begin{enumerate}
\item $ K_{\Lambda+x} = \upsilon_x K_\Lambda $ for all $x \in \bZ^d$
and for all $\Lambda \in \calP_f (\bZ^d)$.
\item $ K_{\Lambda \cup \Lambda'} = K_\Lambda + K_{\Lambda'}$ if
$\Lambda$ and $\Lambda$ are at distance greater than $R$.
\end{enumerate}
The kind of example that we have in mind, and that covers most
applications, is $K_\Lambda = \sum_{X+x\in \Lambda} \upsilon_x A$, for
a given self-adjoint $A \in \calO_X$ (which could be, say,
the magnetization or the occupation operator at the origin, or the
energy in a finite region, or so).

Given a finite-range observable $K$, we can recursively define a
finite-range interaction $\Psi \in \calB^{(f)}$ by means of the
equalities $K_\Lambda \, = \, \sum_{X \subset \Lambda} \psi_X\,$.  We
have a one-to-one correspondence between finite-range macroscopic
observables and finite range interactions.  We can and will consider
more general macroscopic observables by replacing condition 2 with the
condition that the interaction $\Psi$, corresponding to $K$, belongs
to some Banach space.

\subsection{Gibbs-KMS states} 

There are several equivalent ways to characterize the equilibrium
states corresponding to an interaction $\Phi$.  These equivalences
certainly hold if $\Phi \in \calB_\lambda$, for some $\lambda >0$
\cite{Si,BrRo}.  A more general result of this type has been proved
recently in \cite{ArMo}, both for spin and fermion systems, for a
nearly optimal class of interactions.

In this paper, the notation $\ltoi$ will always mean that we take the
limit along an increasing sequence of hypercubes $\Lambda$. 
All our results can presumably also be proved for more general
sequences (Van-Hove limits), but, for simplicity, we will refrain from
doing so.

We denote by $P(\Phi)$ the pressure for the interaction $\Phi$, given
by the limit $P(\Phi)\,=\, \lim_\ltoi |\Lambda|^{-1} \trace (
e^{-H_\Lambda} )$. Here $\trace$ is the normalized trace in
$\calH_\Lambda$ and $H_\Lambda$ is specified by (\ref{H_L}). Let
$\omega$ be a translation invariant state. The mean energy relative to
$\omega$ is defined as $e^{(\Phi)}(\omega) = \overline{H} = \lim_\ltoi
{|\Lambda|^{-1}} \omega( H_\Lambda) $.  Denoting by $\omega_\Lambda$
the restriction of $\omega$ to $\calO_\Lambda$, we define the mean
entropy in the state $\omega$ by $s(\omega)\,=\, \lim_\ltoi
{|\Lambda|^{-1}} S(\omega_\Lambda) $, where $ S(\omega_\Lambda) =
\omega_\Lambda( \log \rho_\Lambda ) = \trace ( \rho_\Lambda \log
\rho_\Lambda )$ and $\rho_\Lambda$ is the density matrix of
$\omega_\Lambda$. The existence of the limits for the pressure, mean
energy and entropy is a standard result.

The variational principle states that 
\begin{equation} \label{varprin}
P(\Phi)\,=\, \sup_{\omega \in \Omega_I} \left( s(\omega)
-e^{(\Phi)}(\omega) \right).
\end{equation}
We denote by $\Omega^{(\Phi)}_I$ the set of states for which the
supremum in Eq.~\eqref{varprin} is attained, and we call such states
the equilibrium states for the interaction $\Phi$.  The set
$\Omega^{(\Phi)}_I$ is a simplex and each of its states has a unique
decomposition into ergodic states.  

The second characterization of equilibrium states is via the KMS
condition.  Let us consider $\tau_t$, a strongly continuous unitary
action of $\bR$ on $\calO$. It is known that, on a norm-dense
subalgebra of $\calO$, $\tau_t$ can be extend to a (pointwise
analytic) action of $\bC$ \cite{BrRo}. So, a state $\omega$ is said to
be $\tau$-KMS if
\begin{equation} \label{kms}
\omega ( A \tau_{i}(B) ) = \omega(BA)
\end{equation}
for all $A$, $B$ in a norm-dense $\tau$-invariant subalgebra of
$\calO$. For a given interaction $\Phi$, one constructs the dynamics
$\tau_t^{(\Phi)}$ as the limit of finite volume dynamics defined, on a
local observable $A$, by $e^{i H_\Lambda t} A e^{-i H_\Lambda
t}$. Then one can speak of a KMS state for the interaction $\Phi$.

The third characterization is through the Gibbs condition. This
condition is analog to the DLR equations for classical spin systems.
Stating it properly would require considerable machinery, including
the Tomita-Takesaki theory. Detailed expositions can be found in
\cite{BrRo,Si} and we will be brief here.  Given an element $P \in
\calO^{(sa)}$ and a state $\omega$, one can define a perturbed state
$\omega^P$ in the following way: Using the Tomita-Takesaki theory one
constructs (in the GNS representation) a dynamics $\tau_t$ that makes
$\omega$ a $\tau$-KMS state. One then perturbs the dynamics $\tau_t$
by formally adding the term $i [P \,,\, {\cdot} \,]$ to its generator
(this would correspond to adding $P$ to the Hamiltonian). Finally, one
defines $\omega^P$ as the KMS state for the perturbed dynamics
(Araki's perturbation theory).

For an interaction $\Phi$, let us consider the perturbation
\begin{equation} \label{W_L}
W_\Lambda \,=\, \sum_{ X \cap \Lambda \ne \emptyset \atop 
X \cap \Lambda^c \ne \emptyset} \phi_X \,,
\end{equation}
which is well-defined under our assumptions. The state $\omega$
satisfies the Gibbs condition if, for every finite subset $\Lambda$,
there exists a state $\omega'$ on $\calO_{\Lambda^c}$ such that
\begin{equation} \label{gibbs}
\omega^{-W_\Lambda} \,=\, \omega^{(\Phi)}_{\Lambda} \otimes \omega'
\,.
\end{equation}
Here $\calO_{\Lambda^c}$ is the subalgebra of observables that ``do
not depend on $\Lambda$'' (we omit the formal definition; suffices to
say that $\calO = \calO_\Lambda \otimes \calO_{\Lambda^c}$). Also,
which is crucial, $\omega^{(\Phi)}_{\Lambda}$ is the finite-volume
Gibbs state on ${\calO}_\Lambda$ given by
\begin{equation}\label{figi}
\omega^{(\Phi)}_\Lambda( A ) \,=\, \frac{\trace( A e^{-H_\Lambda} )}{
\trace( e^{-H_\Lambda} )} \,,
\end{equation} 
The Gibbs condition is very similar to the DLR equations in classical
lattice systems, and it is not difficult to check that the DLR
equations and the Gibbs condition are indeed equivalent for classical
spin systems.

Nor is it hard to verify that finite-volume Gibbs states satisfy all
the previous three conditions.  A fundamental result of quantum
statistical mechanics, due to Lanford, Robinson, Ruelle and Araki,
asserts that the three characterizations are indeed equivalent for
infinite-volume translation invariant states of spins or fermions. The
key to the proof is the Gibbs condition, introduced by Araki. In the
very recent \cite{ArMo}, equivalence has been proved for a very large
class of interactions, much larger than the one considered in this
paper.

\section{Moment generating function} \label{qmgf} 

Given an interaction $\Phi$ with a corresponding Gibbs-KMS state
$\omega \in \Omega^{(\Phi)}_{I}$, and a macroscopic observable $\{
K_\Lambda \}$, uniquely determined by the interaction $\Psi$, we
introduce the moment generating function
\begin{equation} \label{mgf}
f^{(\Psi, \Phi)}(\alpha) \,=\, \lim_\ltoi \frac{1}{|\Lambda|} \log
\omega ( e^{ \alpha K_\Lambda}) \,;
\end{equation} 
that is, when the limit exists. A priori it is not obvious that
$f^{(\Psi, \Phi)}(\alpha)$ depends only on $\Phi$ and not the choice
of $\omega \in \Omega^{(\Phi)}_{I}$. In this paper, however, we will
always work in the one-phase regime, see Remark \ref{rmkkms}.
Furthermore one expects that, as in the classical case, $f^{(\Psi,
\Phi)}(\alpha)$ would depend only on $\Phi$.

We will make one of the following assumptions.
\begin{itemize}
\item[\bf H1:] {\bf High temperature.} Both $\Phi$ and $\Psi$ belong
to some $\calB_\lambda$ and
\begin{equation}
\frac{\lambda}{4} \|\Phi\|_\lambda < 1 \,.
\end{equation}

\item[\bf H2:] {\bf High temperature improved.} $\Phi$ is the sum of
two interactions, $\Phi = \Phi' + \Phi''$, where $\Phi'' =
\{\phi''_x\}_{x \in \bZ^d}$ involves only observables depending on one
site, and, for all $\Lambda \subset \bZ^d$, we have $\left[ H'_\Lambda
\,,\, H''_\Lambda \right] =0$. Also, we assume that $\Phi'$ and $\Psi$
belong to some $\calB_\lambda$ with
\begin{equation}
\frac{\lambda}{4} \|\Phi'\|_\lambda < 1 \,.
\end{equation}
No smallness assumption on $\Phi''$ is made. 

\item[\bf H3:] {\bf Dimension one.} The lattice has dimension one and
both $\Phi$ and $\Psi$ have finite range $R$.
\end{itemize}

\begin{remark}{\rm   
Condition ${\bf H2}$ is important in physical applications where
$\Phi''$ is a chemical potential or an external magnetic field. It
allows us to prove our results at high temperature for {\em any} value
of the chemical potential/magnetic field (see the example in the
introduction).  }\end{remark}

Our main result is

\begin{theorem}\label{main} Let $\omega$, $\Phi$, $\Psi$ be as above.

\begin{enumerate}
\item {\bf (High temperature)} If ${\bf H1}$ or ${\bf H2}$ is
satisfied, then the moment generating function $f^{(\Psi,\Phi)}
(\alpha)$ exists and is finite for all real $\alpha$ such that
\begin{equation}
|\alpha| < \frac{4}{\lambda \|\Psi\|_\lambda}\,.
\end{equation}
If the macroscopic observable is the sum of observables depending only
on one site, i.e., $K_\Lambda= \sum_{x \in \Lambda}\psi_x$, with
$\psi_x \in \calO_{ \{x\} }$, then $f^{(\Psi,\Phi)}(\alpha)$ exists
and is finite for all $\alpha \in \bR$.

\item {\bf (Dimension one)} If ${\bf H3}$ is satisfied, then
$f^{(\Psi,\Phi)}(\alpha)$ exists and is finite for all $\alpha \in
\bR$.
\end{enumerate}

The moment generating function $f^{(\Psi,\Phi)}(\alpha)$ is convex and
Lipschitz continuous; more precisely, 
\begin{equation} \label{lip}
\left| f^{(\Psi,\Phi)}(\alpha_1) - f^{(\Psi,\Phi)}(\alpha_2) \right|
\,\le \, \|\Psi\|_0 | \alpha_1 - \alpha_2| \,,
\end{equation}     
where $\|\Psi\|_0 = \sum_{X \ni 0} \|\psi_X\|$). Moreover

\begin{equation} \label{formula}
f^{(\Psi,\Phi)}(\alpha) \,=\, \lim_\ltoi \frac{1}{|\Lambda|} \log
\left( \frac{\trace(e^{\alpha K_\Lambda} \, e^{-H_\Lambda })}
{\trace(e^{ - H_\Lambda })} \right)\,.
\end{equation}
\end{theorem}    	
    
\begin{remark} \label{rmkkms} {\rm 
Although our proof does not directly use this fact, the assumptions of
Theorem \ref{main} imply that there is a unique KMS state (in
\cite{BrRo}, for instance, check Theorem 6.2.45 for {\bf H1}, Theorem
6.2.46 for {\bf H2}, and Theorem 6.2.47 for {\bf H3}). }
\end{remark} 

\begin{remark} {\rm
The equality of the two limits (\ref{mgf}) and (\ref{formula}) implies
that---using the terminology of \cite{LeLeSp}---semi-local large
deviations are the same as global large deviations. In other words,
$\omega( {\bf 1}_B (|\Lambda|^{-1} K_\Lambda) )$ decreases at the same
exponential rate as $\omega_\Lambda ( {\bf 1}_B (|\Lambda|^{-1}
K_\Lambda) )$.  Global large deviations are so named because they
gauge the probability of deviation from the expected value when a
microscopic observable is averaged over {\em all} the available
volume.  }
\end{remark}

For particular, physically important observables, the results of
Theorem \ref{main} can be improved.

\begin{corollary} 
Suppose that, for all $\Lambda \in \calP_f(\bZ^d)$, the observable
$K_\Lambda$ commutes with the energy $H_\Lambda$.

\begin{enumerate}    
\item If ${\bf H1}$ or ${\bf H2}$ holds, then
$f^{(\Psi,\Phi)}(\alpha)$ exists and is finite for all $\alpha \in
\bR$, and is $C^1$ in a neighborhood of $0$.  If $K_\Lambda$ is the
sum of observables depending only on one site, then
$f^{(\Psi,\Phi)}(\alpha)$ is $C^1$ for all $\alpha$.

\item If ${\bf H3}$ holds, then $f^{(\Psi,\Phi)}(\alpha)$ exists, is
finite, and is $C^1$ for all $\alpha \in \bR$.
\end{enumerate}
\end{corollary}    	

\proof If $\left[ H_\Lambda \,,\, K_\Lambda \right]=0$ then, by
Theorem \ref{main} and Eq.~\eqref{formula},
\begin{eqnarray}
f^{(\Psi,\Phi)}(\alpha) &=& \lim_\ltoi \frac{1}{|\Lambda|} \log
\left( \frac{\trace(e^{\alpha K_\Lambda - H_\Lambda})} {\trace(e^{ -
H_\Lambda })} \right) \nn \\ 
&=& P(\Phi - \alpha \Psi) - P(\Phi) \,,
\end{eqnarray}
so that, as in the classical case, $f^{(\Psi,\Phi)}(\alpha)$ is the
translated pressure. There is a unique Gibbs-KMS state for the
interaction $\Phi - \alpha \Psi$, provided $\|\Phi - \alpha
\Psi\|_\lambda$ is sufficiently small (\cite{BrRo}, Theorem 6.2.45),
so, by the equivalence between Gibbs-KMS states and functionals
tangent to the pressure \cite{Is, Si}, $f^{(\Psi,\Phi)}(\alpha)$ is
differentiable if $\alpha$ is sufficiently small.  If the interaction
$\Psi$ consists only of observables depending on one site, and
$H_\Lambda$ commutes with $K_\Lambda$, then there is a unique
Gibbs-KMS state for $\Phi - \alpha \Psi$, for all $\alpha$, provided
$\|\Phi\|_\lambda$ is small (\cite{BrRo}, Theorem 6.2.46).  If
condition ${\bf H2}$ is satisfied, similar considerations apply 
(see \cite{BrRo}, Theorem 6.2.46). If condition  ${\bf H3}$ 
is satisfied there is a unique Gibbs-KMS state for $\Phi - \alpha \Psi$ 
(\cite{BrRo}, Theorem 6.2.47). \qed

The proof of Theorem \ref{main} is in two steps. In the first step,
instead of $f^{(\Psi,\Phi)}(\alpha)$, we consider
\begin{equation}
g^{(\Psi,\Phi)}(\alpha)\,=\, \lim_\ltoi \frac{1}{|\Lambda|} \log
\trace(e^{\alpha K_\Lambda} e^{-H_\Lambda }) \,.
\end{equation}
In the second step we show that 
\begin{equation}\label{some}
f^{(\Psi,\Phi)}(\alpha)\,=\, g^{(\Psi,\Phi)}(\alpha) - P(\Phi)\,.
\end{equation}
The function $g^{(\Psi,\Phi)}(\alpha)$ is defined via
finite-dimensional objects.  We will prove the existence of the limit
using a subaddivity argument, as in the proof of the existence of the
pressure. The equality \eqref{some} is proved using perturbation
theory for KMS states.

\subsection{Perturbation of KMS states}

A basic ingredient in the proof of the existence of the pressure is
the following matrix inequality:
\begin{equation}\label{b1}
\left| \log \trace \left( e^{H+P} \right) - \log \trace \left( e^{H}
\right) \right| \,\le\, \| P \| \,,
\end{equation}
where $H$ and $P$ are symmetric $n \times n$ matrices.  In order to
study the function $g^{(\Psi,\Phi)}(\alpha)$, where we have two
(generally non-commuting) exponentials under the trace, one needs to
estimate quantities like
\begin{equation}\label{b2}
\left| \log \trace \left( C e^{H+P} \right) - \log \trace \left( C
e^{H} \right) \right| \,,
\end{equation}
where $C$ is a positive-definite $n \times n$ matrix.  A little
thinking convinces one that an estimate of \eqref{b2} by a constant
times $\|P\|$ cannot possibly hold true, if the constant is required
not to depend on $C$ or $n$.

The following lemma gives an upper bound for \eqref{b2} which is
independent of $C$ and $n$, although it has a different form than
Eq.~\eqref{b1}.

\begin{lemma} \label{matrix} 
Let $H, P \in \bC^{n \times n}$, with $H^* = H$ and $P^* = P$. 
\begin{enumerate}
\item We have 
\begin{equation}
\left| \log \trace \left( e^{H+P} \right) - \log \trace \left( e^{H}
\right) \right| \,\le\, \| P \| \,.
\end{equation}
\item  Also, if $C \in \bC^{n \times n}$ with $C > 0$,
\begin{equation}
\left| \log \trace \left( C e^{H+P} \right) - \log \trace \left( C
e^{H} \right) \right| \,\le\, \sup_{0 \le t \le 1} \, \sup_{-\frac12
\le s \le \frac12} \left\| U^{-s}(t) \,P\, U^{s}(t) \right\| \,,
\end{equation}
where
\begin{equation}
U^s(t) = e^{s( H+t P)} \,.
\end{equation}
\end{enumerate}
\end{lemma}

\proof The proof of part 1 is standard. One writes 
\begin{eqnarray}
&&\bi \left| \log \trace \left( e^{H+P} \right) - \log \trace \left(
e^{H} \right) \right| \,=\, \left| \int_0^1\, dt \, \frac{d}{dt} \log
\trace \left( e^{H+tP} \right) \right| \nn \\
&\le& \int_0^1\, ds \left| \frac{ \trace ( P e^{H+tP} ) } {\trace (
e^{H+tP} ) } \right| \,\le\, \|P\| \,,
\end{eqnarray}
having used the fact that, for $E \ge 0$,
\begin{equation}\label{b3}
\left| \frac{\trace(A E)}{\trace(E)} \right| \le \|A\| \,.
\end{equation}

To prove part 2, we recall DuHamel's identity for the derivative of
$e^{F(t)}$, when $F(t)$ is a bounded operator:
\begin{equation} \label{duha}
\frac{d}{dt} e^{F(t)} \,=\, \int_0^1 du \, e^{uF(t)} \, F'(t) \,
e^{(1-u)F(t)} \,.
\end{equation}
We write
\begin{equation}
\log \trace \left( C e^{H+P} \right) - \log \trace \left(C e^{H}
\right) \,=\, \int_0^1 dt \, \frac{d}{dt} \log \trace \left( C e^{H +
tP} \right)
\end{equation}
and 
\begin{eqnarray}
&&\bi \frac{d}{dt} \log \trace \left( C \, e^{H + tP} \right) \nn \\ 
&=& \frac{ \trace\left( \int_0^1 du \, C \, e^{u(H+tP)} \, P \,
e^{(1-u)(H+tP)} \right)} {\trace\left( C \, e^{H+tP} \right)} \nn \\
&=& \frac{ \trace\left( e^{(H+tP)/2} \, C \, e^{(H+tP)/2} \int_0^1 du
\, e^{(u-1/2)(H+tP)} \, P \, e^{(1/2-u)(H+tP)} \right)} {\trace\left(
e^{(H+tP)/2} \, C \, e^{(H+tP)/2} \right)} \nn \\
&\le& \left\| \int_{-1/2}^{1/2} ds \, e^{-s(H+tP)} \, C \, e^{s(H+tP)}
\right\| \,,
\end{eqnarray}
where we have used the bound \eqref{b3} with $E= e^{(H + tP)/2} C
e^{(H+tP)/2}$.  This concludes the proof of Lemma \ref{matrix}. \qed

Lemma \ref{matrix} involves the quantity $U^{-s}(t) P U^{s}(t)$, which
is the time evolution (in imaginary time) of the observable $P$,
relative to the dynamics generated by $H+tP$.  One needs to estimate
the dynamics for imaginary times between $-i/2$ and $i/2$. The
connection with the KMS boundary conditions is evident.

If we define a (finite-volume) state $\omega$ and a perturbed state
$\omega^P$ by
\begin{equation}
\omega(A) \,=\, \frac{\trace(A e^{H})}{\trace(e^{H})} \,, \qquad
\omega^P(A) \,=\, \frac{ \trace ( A e^{H+P} )}{\trace ( e^{H+P}) } \,,
\end{equation}
then Lemma \ref{matrix} immediately implies that, for $C > 0$,
\begin{equation}
\left|\log \omega^P(C) - \log \omega(C) \right| \,\le \, \|P\| +
\sup_{0 \le t \le 1} \, \sup_{-\frac12 \le s \le \frac12} \|
U^{-s}(t) \,P\, U^{s}(t) \| \,.
\end{equation}
We will generalize this bound for Gibbs-KMS states of the infinite
system, using results from the perturbation theory of KMS states (see,
e.g., Chapter 5.4 of \cite{BrRo} or Chapter IV.5 of \cite{Si}).  For a
$\tau$-KMS state $\omega$, we denote by $( \calG_\omega, \pi_\omega,
O_\omega ) = ( \calG , \pi, O )$ its GNS representation.  The scalar
product on $\calG$ is indicated with $\la \, \cdot \,,\, \cdot \,
\ra$. For any $A\in \calO$ we have
\begin{equation}
\omega(A) \,=\, \la O, \pi(A) O \ra 
\end{equation} 
and the dynamics $\tau$ is implemented by some self-adjoint operator
$H$ on $\calG$:
\begin{equation}
\pi(\tau_s(A)) \,=\, e^{is H} \pi(A) \, e^{-is H}\,.
\end{equation}
From now on we will identify an element $A$ with its representative
$\pi(A)$. This is possible since the two-sided ideal $\{ A \in \calO
\,|\, \omega(A^* A) = 0 \}$ is trivial (\cite{Si}, Theorem IV.4.10),
therefore $\pi$ is the left multiplication on $\calO$ (\cite{Si},
Theorem I.7.5). 

For $P \in \calO^{(sa)}$, $\tau^{P}(A)$  given by 
\begin{eqnarray}
&& \!\! \tau_s^P(A) = \tau_s(A) \nn \\ 
&& + \sum_{n=1}^\infty i^{n} \int_0^s ds_1 \int_0^{s_1} ds_2 \cdots
\int_{0}^{s_{n-1}} \!\! ds_n \left[ \tau_{s_n}(P), \left[ \cdots [
\tau_{s_1}(P) , \tau_{s}(A) ] \right] \cdots \right] \,. \nn \\
\label{taup}
\end{eqnarray}
defines a strongly continuous semigroup of automorphims of $\cal O$
implemented by $H+P$:
\begin{equation}
\tau^{P}_{s}(A) \,=\, e^{ is (H +P)} A \, e^{-is (H+P)} \,.
\end{equation}
Moreover we have 
\begin{equation} \label{perdyn}
\tau^{P}_{s} (A) \,=\, \Gamma_{s}^{P} \tau_{s}(A) (\Gamma_{s}^{P})^*
\,=\, \Gamma_{s}^{P} \tau_{s}(A) (\Gamma_{s}^{P})^{-1} \,,
\end{equation}
where the unitary operator 
\begin{equation} \label{unit}
\Gamma_{s}^{P}\,=\, e^{ is (H+P)}  e^{-is H} 
\end{equation}
has the following representation as norm-convergent series:
\begin{equation} \label{gammas}
\Gamma_{s}^{P} = {\bf 1} + \sum_{n=1}^\infty i^{n} \int_0^s ds_1
\int_0^{s_1} \! ds_2 \cdots \int_{0}^{s_{n-1}} \! ds_n \,
\tau_{s_n}(P) \cdots \tau_{s_1}(P) \,. 
\end{equation}
Furthermore, $f(s,P) = \Gamma_s^P O$, defined on $\bR$, extends to a
holomorphic function $f(z,P)$ on $\{ z\in \bC \,|\, 0 \le {\rm Im}z
\le 1/2 \}$ (i.e., the function is continuous and bounded on the close
strip, and analytic on its interior). In particular, $O$ belongs to
the (maximal) domain of $\Gamma_{i/2}^P$, so that one can set
\begin{equation}
O^P \,=\, \Gamma_{i/2}^P O \,=\, e^{-(H+P)/2}\, e^{H/2} \, O\,.
\end{equation}
Araki's perturbation theory asserts that the state 
$\omega^P$ given by
\begin{equation} \label{pert}
\omega^P (A) \,=\, \frac{ \la O^P \,,\, A \, O^P \ra} {\la O^P \,,\,
O^P \ra } \,=\, \frac{ \left\la O \,,\, (\Gamma^P_{i/2})^* A \,
(\Gamma^P_{i/2}) \, O \right\ra } { \left\la O \,,\,
(\Gamma^P_{i/2})^* (\Gamma^P_{i/2}) \, O \right\ra }
\end{equation}
is a $\tau^P$-KMS state.  

The bound in Lemma \ref{matrix} involves the norm of the
imaginary-time evolution of the perturbation $P$. Therefore, for
infinite systems, we will assume that $P$ is an analytic element for
the dynamics in the strip $\{ |{\rm Im}z| \le 1/2 \}$: by this mean
that $\tau_z(P)$ extends to a holomorphic function in the strip, in
the sense specified above. This is clearly a strong assumption and the
main limitation of our approach.

\begin{theorem}  \label{infmatrix}
Let $\omega$ be a $\tau$-KMS state and let $P \in \calO$ be a
self-adjoint analytic element in the strip $\{ |{\rm Im}z| \le 1/2
\}$.  Then, for all positive $C \in \calO$ we have
\begin{equation} \label{sds}
\left| \log \omega^{P}(C) -\log \omega(C) \right| \,\le \, \|P\| +
\sup_{0\le t \le 1} \, \sup_{-\frac12 \le s \le \frac12} \|
\tau^{tP}_{is}(P)\|
\end{equation}
\end{theorem}

\proof The proof of Theorem \ref{infmatrix} follows closely the proof
of Lemma \ref{matrix}.  We first assume that $C^{1/2}$ is an analytic
element for the dynamics $\tau$---such elements form a dense
subalgebra of $\calO$ (\cite{Si}, Proposition IV.4.6). Rewriting
Eq.~\eqref{gammas} as
\begin{equation} \label{gammas1}
\Gamma_{s}^{P} \,=\, {\bf 1} + \sum_{n=1}^\infty (is)^{n} \int_0^1
du_1 \int_0^{u_1} \! du_2 \cdots \int_0^{u_{n-1}} \! du_n \, \tau_{s
u_n}(P) \cdots \tau_{s u_1}(P)
\end{equation}
and recalling the hypothesis on $P$, it is easy to extend
$\Gamma_{s}^{P}$ to a holomorphic function on $\{ |{\rm Im}s| \le 1/2
\}$. In light of Eq.~\eqref{perdyn}, then, we conclude that $C^{1/2}$
is an analytic element for $\tau^{tP}$ in that same strip, for all $0
\le t \le 1$.

Using Eq.~\eqref{pert} we have 
\begin{eqnarray}
&&\bi \log \omega^P(C) - \log \omega(C) \,=\, \int_0^1 dt \,
\frac{d}{dt} \log \omega^{tP}(C) \nn \\
&=& \int_0^1 dt \, \frac{d}{dt} \left[ \log \left\la O \,,\,
(\Gamma^{tP}_{i/2})^* \, C \, \Gamma^{tP}_{i/2} \, O \right\ra - \log
\left\la O^{tP} \,,\, O^{tP} \right\ra \right] \,.
\label{ma3}
\end{eqnarray}
We now claim that
\begin{equation} \label{ma1}
\frac{d}{dt} \Gamma^{tP}_{i/2} \,=\, - \int_0^{1/2} ds\,
\tau_{is}^{tP} (P) \, \Gamma^{tP}_{i/2} \,.
\end{equation}
Verifying \eqref{ma1} would amount to a simple application of
DuHamel's formula \eqref{duha}, if $H$ were a bounded operator. In the
case at hand we need to work a little harder, even though we use the
same idea. For $e>0$, let $\Pi_e$ be the projection on the invariant
space of $H$ defined by values of its spectral measure in $[-e,
e]$. Then $\Pi'_e = {\bf 1} - \Pi_e$ is the projection on the
orthogonal space. Set
\begin{equation}
H_e = \Pi_e \, H \, \Pi_e\,, \qquad H'_e = \Pi'_e \, H \, \Pi'_e \,,
\qquad P_e = \Pi_e \, P \, \Pi_e \,.
\end{equation}
Clearly, $H_e$ and $P_e$ are bounded operators and $[H'_e \,,\, H_e] =
[H'_e \,,\, P_e] = 0$.  By means of \eqref{duha}, and after a change
of variable, we verify that
\begin{equation} 
\frac{d}{dt} e^{-(H_e+tP_e)/2} e^{H_e/2} = - \int_0^{1/2} \!
ds\, e^{-s(H_e+tP_e)} P_e e^{s(H_e+tP_e)} e^{-(H_e+tP_e)/2} \,
e^{H_e/2} \,.
\end{equation}
Now we multiply each factor above by the corresponding term
$e^{uH'_e}$ ($u = \pm 1/2, \pm s$); these terms commute with
everything. We obtain
\begin{equation} \label{ma15}
\frac{d}{dt} \Gamma^{tP_e}_{i/2} \,=\, - \int_0^{1/2} ds\,
\tau_{is}^{tP_e} (P_e) \, \Gamma^{tP_e}_{i/2} \,.
\end{equation}
That \eqref{ma15} becomes \eqref{ma1}, as $e \to +\infty$, follows
from \eqref{taup}---or rather its analytic continuation---and
\eqref{gammas1}, since $P_e$ is entire analytic for $\tau_s$, and $\|
P_e - P \| \to 0$.

Once \eqref{ma1} is settled, we can write
\begin{eqnarray}
&&\bi \frac{d}{dt} \log \left\la O \,,\, (\Gamma^{tP}_{i/2})^* \, C \,
\Gamma^{tP}_{i/2} \, O \right\ra \nn \\
&=& - \, \frac{ \omega^{tP} \left( \int_{-1/2}^{0} ds \,
\tau_{is}^{tP}(P) C + \int_{0}^{1/2} ds\, C \tau_{is}^{tP}(P) \right)
} {\omega^{tP} (C) } \,.
\label{eq34}
\end{eqnarray}
The symmetric form of the KMS condition for $\omega^{tP}$ is easily
derived from \eqref{kms}: for $A, B$ analytic in the strip, 
\begin{equation}
\omega^{tP}\left( \tau^{tP}_{-i/2}(A) \, \tau^{tP}_{i/2}(B)\right)
\,=\, \omega^{tP}\left( B A\right) \,.
\end{equation}
Applying the above twice,
\begin{eqnarray}
\omega^{tP} \left( \tau_{is}^{tP}(P) C \right) &=& \omega^{tP} \left(
\tau^{tP}_{-i/2}(C^{1/2}) \, \tau_{i(s+1/2)}^{tP}(P) \,
\tau_{i/2}^{tP}(C^{1/2}) \right) \,; \nn \\
\omega^{tP} \left( C \tau_{is}^{tP}(P) \right) &=& \omega^{tP} \left(
\tau^{tP}_{-i/2}(C^{1/2}) \, \tau_{i(s-1/2)}^{tP}(P) \,
\tau_{i/2}^{tP}(C^{1/2}) \right) \,.
\end{eqnarray}
We thus turn \eqref{eq34} into
\begin{eqnarray}
&&\bi \frac{d}{dt} \log \left\la O^{tP} \,,\, C \, O^{tP}
\right\ra \nn \\
&=& -\, \frac{ \omega^{tP} \left( \tau^{tP}_{-i/2}(C^{1/2}) \,
\int_{-1/2}^{1/2} ds\, \tau_{is}^{tP}(P) \, \tau_{i/2}^{tP}(C^{1/2})
\right) } {\omega^{tP} \left( \tau^{tP}_{-i/2}(C^{1/2})
\tau^{tP}_{i/2}(C^{1/2}) \right)} \,,
\label{eq35}
\end{eqnarray}
and therefore
\begin{equation} \label{ma2}
\left| \frac{d}{dt} \log \left\la O^{tP} \,,\, C \, O^{tP} \right\ra
\right| \le \sup_{-\frac12 \le s \le \frac12} \left\|
\tau_{is}^{tP}(P) \right\| \,.
\end{equation}
Here we have used the fact that
\begin{equation}
A \mapsto \frac{\omega( B^* A B)}{\omega(B^* B) }
\end{equation}
defines a state on $\calO$ if $\omega(B^* B) \ne 0$. 

As for the second term in \eqref{ma3}, we plug $C = {\bf 1}$ in
\eqref{eq35}, use the invariance of $\omega^{tP}$ with respect to
$\tau_z^{tP}$, and conclude that
\begin{equation}
\left| \frac{d}{dt} \log \left\la O^{tP} \,,\, O^{tP} \right\ra
\right| \,\le\, \|P\|\,.
\end{equation}
This gives the desired bound when $C^{1/2}$ is analytic. The general
statement follows by density, see Corollary IV.4.4 in \cite {Si}. \qed

\subsection{Analyticity estimates} \label{analyticity}

As is apparent from the previous section, we need estimates on the
evolution of observables (in imaginary time). We will use two results,
one valid at high temperature and one valid in dimension one.

The first is due to Ruelle, has no restriction on the dimension, and
is a standard.

\begin{prop} \label{anal1}
Let $\Phi \in \calB_\lambda$, for some $\lambda>0$ (see Section
\ref{interactions}). For any $\Lambda \in \calP_f(\bZ^d)$ and any
collection of numbers $\{ u_X \}_{X \subset \Lambda}$, with $u_X =
u_X(\Lambda) \in [0,1]$, set
\begin{equation} \label{ma4}
H_\Lambda^{(u)} = \sum_{X \subset \Lambda} u_X \phi_X
\end{equation}
(of course, $H_\Lambda^{(u)} = H_\Lambda$, if $u_X =1$ for all $X$).
If $A \in \bigcup_X \calO_X$ is a local observable and $z$ belongs to
the strip $\{ |{\rm Im}z| \le 2/(\lambda \|\Phi\|_\lambda) \}$, then
\begin{equation}
\left\| e^{iz H_\Lambda^{(u)}} A e^{-iz H_\Lambda^{(u)}} \right\|
\,\le\, \frac{1}{ 1 - |{\rm Im}z| \, \frac{\lambda}{2} \,
\|\Phi\|_\lambda} \, \|A\| \, e^{\lambda|X|} \,.
\end{equation}
This estimate is uniform in $\Lambda$ (and $\{ u_X \}$) and thus holds
in the limit $\ltoi$, when this limit exists. In particular it holds
for the infinite-volume dynamics $\tau_z$.
\end{prop}

\proof Follows trivially from the estimates of Theorem 6.2.4 in
\cite{BrRo}.
\qed

Theorem \ref{anal1} implies that, in the high-temperature regime
\begin{equation} 
\frac{\lambda}{4}\|\Phi\|_\lambda < 1 \,,
\end{equation}
local observables are analytic elements for the dynamics at least in
the strip $\{ |{\rm Im}z| \le 1/2 \}$, which is what we need.
\medskip

The second estimate is due to Araki \cite{Ar1} and applies only in
dimension one.  It was used recently in \cite{Ma1} to prove a central
limit theorem in one-dimensional spin systems. 

In order to state it we introduce the concept of exponentially
localized observables. Denote $\calO_n = \calO_{[-n,n]}$.  Given
$A \in \calO$, we set $\|A\|^{[0]}=\|A\|$ and
\begin{equation}
\|A\|^{[n]} \,=\, \inf_{A_n \in \calO_n} \|A-A_n\| \,.
\end{equation}
This allows us to define, for $0 < \theta < 1$, the norm
\begin{equation}
\|A\|_{(\theta)} \,=\, \sum_{n \ge 0} \theta^{-n} \, \|A\|^{[n]} \,.
\end{equation}
An element $A$ of $\calO$ is said to be exponentially localized with
rate $\theta$ if, and only if, $\|A\|_{(\theta)} < \infty$. The symbol
$\calO^{(\theta)}$ will denote the space of all such observables.

We consider an interaction $\Phi$ of finite range $R$, and set
\begin{equation}
S( \Phi) \,=\, \left\| \sum_{X \ni 0} \frac{\phi_X}{\diam(X)} \right\|
\,.
\end{equation}
Also, for $s>0$, we define
\begin{equation}
F_R(s) \,=\, \exp \left[ (-R+1)s + 2 \sum_{k=1}^R \frac{e^{kR}-1} {k}
\right] \,.
\end{equation}
We have 

\begin{prop} \label{anal2} 
Let $\Phi \in \calB^{(f)}$, with range $R$. If $\theta \in (0,1)$ and
$h >0$ verify $\theta e^{4h S(\Phi)} = \theta' < 1$, then there exists
a constant $M=M(R,\theta,h)$ (independent of $\Phi$) such that, for $A
\in \calO^{(\theta)}$ and $|{\rm Im}z| \le h$,
\begin{equation}
\left\| e^{iz H_\Lambda^{(u)}} A e^{-iz H_\Lambda^{(u)}}
\right\|_{(\theta')} \le \, M\, F_R( 2 S(\Phi)) \, \|A\|_{(\theta)}
\,.
\end{equation} 
Here $H_\Lambda^{(u)}$ is defined as in \eqref{ma4}. This estimate is
uniform in $\Lambda$ (and $\{ u_X \}$) and thus holds in the limit
$\ltoi$, when this limit exists. In particular it holds for the
infinite-volume dynamics $\tau_z$.
\end{prop}

\proof Follows from the results of \cite{Ar1}; see also \cite{Ma2}. 
\qed

We will use this result in the particular case in which the
macroscopic observable $\{ K_\Lambda \}$ has finite range. Hence
notice that, if $A$ is a local observable, then $A \in
\bigcap_{\theta} \calO^{(\theta)}$. Furthermore, for every $\theta \in
(0,1)$, there exists a constant $D=D(\theta,R')$ such that
$\|A\|_{(\theta)} \,\le\, D \|A\|$, for all $A \in \calO_X$ with
$\diam(X) \le R'$. The reverse bound, $\|A\| \le \|A\|_{(\theta)}$, is
of course valid for every $A \in \calO$.  These considerations and
Proposition \ref{anal2} imply that, for any such $A$, there exists a
constant $G=G(R,R',S(\Phi))$ such that, for $|{\rm Im}z| \le 1/2$,
\begin{equation} \label{ma9}
\left\| e^{iz H_\Lambda} A e^{-iz H_\Lambda} \right\| \,\le\,  
G \, \|A\| \,.
\end{equation} 
Once again, this is what we need to apply Lemma \ref{matrix} and
Theorem \ref{infmatrix}.

\subsection{Subadditivity}

We now give sufficient conditions for the limit
\begin{equation} \label{gal}
g^{(\Psi,\Phi)}(\alpha) \,=\, \lim_\ltoi  \frac{1}{|\Lambda|} 
\log \trace(e^{\alpha K_\Lambda} e^{-H_\Lambda })
\end{equation}
to exist. 

\begin{theorem} \label{likepressure} 
The following holds true:
\begin{enumerate} 
\item {\bf (High temperature)} If condition {\bf H1} or {\bf H2}
applies, then the function $g^{(\Psi,\Phi)}(\alpha)$ defined by
Eq.~\eqref{gal} exists and is finite for $\alpha$ real, with
\begin{equation}
|\alpha| < \frac{4}{\lambda \|\Psi\|_\lambda} \,.
\end{equation}
Furthermore, if $\Psi=\{\psi_x\}_{x\in \bZ^d}$ with $\psi_x \in
\calO_{ \{x\} }$, (observables depending only on one site), then
$g^{(\Psi,\Phi)}(\alpha)$ exists and is finite for all $\alpha \in
\bR$.

\item {\bf (Dimension one)} If condition {\bf H3} applies, then
$g^{(\Psi,\Phi)}(\alpha)$ exists and is finite for all $\alpha \in
\bR$.
\end{enumerate}
\end{theorem}

\proof We start with item 1 under the condition {\bf H1}. The proof
combines Lemma \ref{matrix}, the analyticity estimates of Section
\ref{analyticity}, and a subaddivity argument as in the proof of the
existence of the pressure.  Let $\Lambda$ to be an hypercube of side
length $L$. We choose $a > 0$ and write $L=na+b$, with $0\le b < a$.
We divide the $L$-cube into disjoint adjacent $n^d$ $a$-cubes,
$\cube_1, \cube_2, \cdots, \cube_{n^d}$ and a ``rest'' region
$\cube_0$ which contains $L^d- (na)^d$ lattice points. We write
\begin{equation}
H_\Lambda \,=\, \sum_{j=1}^{n^d} H_{\cube_j} + H_{\cube_0} + W\,,
\qquad K_\Lambda \,=\, \sum_{j=1}^{n^d} K_{\cube_j} + K_{\cube_0} + U
\,.
\end{equation}
where
\begin{equation} 
W \,=\, \sum_{X} {}^{'} \phi_X \,, \qquad U \,=\, \sum_{X} {}^{'}
\psi_X \,
\end{equation}
and $\sum^{'}$ indicates a sum over all $X \subset \Lambda$ such that,
for some $j = 0, 1, \dots, n^d$, $X \cap {\cube_j} \not= \emptyset$
and $X \cap {\cube_j^c} \not= \emptyset$.  We denote by
\begin{equation}
g_\Lambda^{(\Psi,\Phi)}(\alpha)\,=\, \frac{1}{|\Lambda|}\log \trace
\left( e^{\alpha K_\Lambda} e^{-\beta H_\Lambda} \right)
\end{equation}
the function whose limit we are set to take. By the commutativity
property of local observables and the translation invariance, 
\begin{eqnarray}
\log \trace \left( e^{\, \alpha \sum_{j=1}^{n^d} K_{\cube_j} -
\sum_{j=1}^{n^d} H_{\cube_j}} \right) &=& \log \prod_{j=1}^{n^d}
\trace \left( e^{\alpha K_{\cube_j}} e^{-\beta H_{\cube_j}} \right)
\nn \\
&=& (na)^d g_{\cube_1}^{(\Psi,\Phi)}(\alpha)\,.
\label{ma5}
\end{eqnarray}
Set now
\begin{equation}
P\,=\,H_{\cube_0} + W \,, \qquad  Q\,=\,K_{\cube_0} + U \,.
\end{equation} 
Using Eq.~\eqref{ma5}, the triangle inequality, Lemma \ref{matrix} and
Proposition \ref{anal1}, we are able to estimate
\begin{eqnarray}
&&\bi \left| g_\Lambda^{(\Psi,\Phi)}(\alpha) - \frac{(na)^d}
{|\Lambda|} g^{(\Psi,\Phi)}_{\cube_1}(\alpha) \right| \nn \\
&\le& \frac{1}{|\Lambda|} \sup_{0 \le t \le 1} \, \sup_{-\frac12 \le s
\le \frac12} \| e^{-s (H_\Lambda -tP)}\, P\, e^{s (H_\Lambda -tP)}\|
\nn \\
&\phantom{\le}& +\: \frac{1}{|\Lambda|} \sup_{0 \le t \le 1} \,
\sup_{-\frac12 \le s \le \frac12} \| e^{-s\alpha (K_\Lambda -tQ)} \,
\alpha T \, e^{s\alpha(K_\Lambda -tQ)} \| \nn \\
&\le& \frac{1}{1 - \frac{\lambda}{4} \|\Phi\|_\lambda} \,
\frac{1}{|\Lambda|} \left(\sum_{X \subset \cube_0} +
\sum_{X}{}^{'}\right) \|\phi_X\| \, e^{\lambda |X|} \nn \\
&\phantom{\le}& +\: \frac{|\alpha|}{1 - |\alpha|
\frac{\lambda}{4}\|\Phi\|_\lambda} \, \frac{1}{|\Lambda|}
\left(\sum_{X \subset \cube_0} + \sum_{X}^{}{'}\right) \|\psi_X\| \,
e^{\lambda|X|} \,.
\label{ineq}
\end{eqnarray}
We take the limit $\ltoi$ of the various parts of Ineq.~\eqref{ineq}.
First,
\begin{eqnarray}
\frac{1}{|\Lambda|}\sum_{X \in \cube_0} \|\phi_X\| \, e^{\lambda |X|}
&\le& \frac{1}{L^d} \sum_{x \in \cube_0} \, \sum_{X \ni x} \|\phi_X \|
\, e^{\lambda |X|} \nn \\
&\le& \frac{L^d - (na)^d}{L^d} \|\Phi\|_\lambda \,\longrightarrow\, 0
\end{eqnarray}
as $L \to \infty$; similarly for $\sum_{X \in \cube_0} \|\psi_X\| \,
e^{\lambda |X|}$. Also, in the same limit,
\begin{eqnarray}
\frac{1}{|\Lambda|} \sum_X {}^{'} \|\phi_X\| \, e^{\lambda |X|} &\le&
\frac{1}{L^d} \sum_{j=1}^{n^d} \, \sum_{X \cap \cube_j \not= \emptyset
\atop X \cap \cube_j^c \not= \emptyset} \|\phi_X\| \, e^{\lambda |X|}
\nn \\
&\le& \frac{n^d}{L^d} \sum_{X \cap \cube_{1} \not= \emptyset \atop X
\cap \cube_{1}^{c} \not= \emptyset } \|\phi_{X}\| \, e^{\lambda |X|}
\nn \\
&\longrightarrow& \frac{1}{|\cube_{1}|} \sum_{X \cap \cube_{1} \not=
\emptyset \atop X \cap \cube_{1}^{c} \not=\emptyset} \|\phi_{X}\| \,
e^{\lambda |X|} \,.
\label{ma6}
\end{eqnarray}
Once again, a similar estimate holds for $\sum_X^{'} \|\psi_X\| \,
e^{\lambda |X|}$. In the remainder, for the sake of the notation, we
rename $\cube_1 = \cube$. From \eqref{ineq}-\eqref{ma6} we obtain
\begin{eqnarray}
&&\bi \left| \limsup_{\ltoi} g^{(\Psi,\Phi)}_\Lambda(\alpha) -
g^{(\Psi,\Phi)}_{\cube} (\alpha) \right| \nn \\
&\le& \frac{1}{1 - \frac{\lambda}{4} \|\Phi\|_\lambda}
\frac{1}{|\cube|} \sum_{X \cap \cube \not= \emptyset \atop X \cap
\cube^{c} \not= \emptyset } \|\phi_{X}\| e^{\lambda |X|} \nn \\
&\phantom{\le}& +\: \frac{|\alpha|}{1 - |\alpha| \frac{\lambda}{4}
\|\Psi\|_\lambda} \frac{1}{|\cube|} \sum_{X \cap \cube \not= \emptyset
\atop X \cap \cube^{c} \not= \emptyset } \|\psi_{x} \| e^{\lambda|X|}
\,.
\label{ma7}
\end{eqnarray}

It is now time to take the limit $\cube \nearrow \bZ^d$. Denote by
$\cube'$ the cube of side length $a - a^{1/2}$ and concentric to
$\cube$. We have
\begin{eqnarray}
&&\bi \frac{1}{|\cube|} \sum_{X \cap \cube \not= \emptyset \atop X
\cap \cube^{c} \not= \emptyset } \|\phi_X\| \, e^{\lambda|X|} \nn \\
&\le& \frac{1}{|\cube|} \sum_{x \in \cube'} \sum_{X \ni x \atop X \cap
\cube^{c} \not= \emptyset} \|\phi_X\| \, e^{\lambda|X|} +
\frac{1}{|\cube|} \sum_{x \in \cube \setminus \cube'} \sum_{X \ni x
\atop X \cap \cube^{c} \not= \emptyset} \|\phi_X\| \, e^{\lambda|X|}
\nn \\
&\le& \frac{|\cube'|}{|\cube|} \sum_{\diam(X) \ge a^{1/2}} \|\phi_X\|
\, e^{\lambda|X|} + \frac{|\cube \setminus \cube'|}{|\cube|} \| \Phi
\|_{\lambda} \,\longrightarrow\, 0 \,,
\label{ma11}
\end{eqnarray}
as $a \to \infty$. The same holds for the second term of \eqref{ma7}.
Finally, then,
\begin{equation}
\left| \limsup_{\ltoi } g^{(\Psi,\Phi)}_\Lambda(\alpha) -
\liminf_{\ltoi } g^{(\Psi,\Phi)}_\Lambda(\alpha) \right| \,=\,0 \,.
\end{equation}
which proves the existence and finiteness of the limit
$g^{(\Psi,\Phi)}(\alpha)$, in the high temperature regime.

In the special case in which $\Psi$ consists only of one-body
interactions, we have
\begin{equation}
K_\Lambda \,=\, \sum_{j=1}^{n^d} K_{\cube_j} +  K_{\cube_0} \,,
\end{equation}
i.e., $U = 0$ and all the observables involved commute.  Thus, in the
first inequality of \eqref{ineq}, the second term simplifies to
\begin{equation} \label{ma8}
\| e^{-s\alpha (K_\Lambda -tK_{\cube_0})} \alpha K_{\cube_0}
e^{s\alpha (K_\Lambda -tK_{\cube_0})} \| = \|\alpha K_{\tilde \cube}\|
\le (L^d -(na)^d) |\alpha| \, \|\Psi\|_0 \,.
\end{equation}
Proceeding as above, one proves the existence of
$g^{(\Psi,\Phi)}(\alpha)$ for all $\alpha\in \bR$.
\medskip

If condition {\bf H2} holds instead of {\bf H1} we have to modify
the argument a little: using the same notation as above and because
$\Phi''$ only involves one-site interactions, we have
\begin{equation} \label{deco}
H'_\Lambda \,=\, \sum_{j=1}^{n^d} H'_{\cube_j} + H'_{\cube_0} + W' \,,
\qquad H''_\Lambda \,=\, \sum_{j=1}^{n^d} H''_{\cube_j} +
H''_{\cube_0} \,,
\end{equation}
We note that since $\left[ H'_V \,,\, H''_V \right] = 0$ for all $V\in
\calP_f(\bZ^d)$, then the decomposition \eqref{deco} implies that
\begin{equation} \label{specom}
\left[H''_V \,,\, W' \right] \,=\,0\,.
\end{equation}  
In order to estimate  
\begin{equation}
\log \trace \left( C \, e^{- H_\Lambda} \right) - \log \trace \left( C
\, e^{- \sum_{j=1}^{n^d} H_{\cube_j}} \right)
\end{equation}
for positive $C$, we proceed in two steps, using Lemma \ref{matrix}. We
have, using \eqref{specom},
\begin{eqnarray}
&&\bi \left| \log \trace \left( C\, e^{- H_\Lambda} \right) -  
\log \trace \left( C\, e^{-H_\Lambda -W'} \right) \right| \nn \\
&\le& \sup_{0 \le t \le 1} \, \sup_{-\frac12 \le s \le \frac12}
\left\| e^{-s (H_\Lambda -tW')} \, W' \, e^{s (H_\Lambda -tW')}
\right\| \nn \\
&=& \sup_{0 \le t \le 1} \, \sup_{-\frac12 \le s \le \frac12} \left\|
e^{-s (H'_\Lambda-tW')} \, e^{-s H''_\Lambda} \, W' \, e^{s
H''_\Lambda} \, e^{s (H'_\Lambda-tW')} \right\| \nn \\
&=& \sup_{0 \le t \le 1} \, \sup_{-\frac12 \le s \le \frac12} \left\|
e^{-s (H'_\Lambda -tW')} \, W' \, e^{s (H'_\Lambda-tW')} \right\| \,.
\end{eqnarray}
This term does not involve $\Phi''$ anymore and is estimated as under
condition ${\bf H1}$.  On the other hand, since $H_\Lambda-W' =
\sum_{j} H_{\cube_j} + H_{\cube_0}$ is a sum of commuting terms, we
have
\begin{eqnarray}
&&\bi \left| \log \trace \left( C \, e^{- (H_\Lambda-W')} \right) -
\log \trace \left( C \, e^{- \sum_{j=1}^{n^d} H_{\cube_j}} \right)
\right| \nn \\
&\le& \sup_{0 \le t \le 1} \, \sup_{-\frac12 \le s \le \frac12}
\left\| e^{-s \left(\sum_{j} H_{\cube_j} -t H_{\cube_0} \right)} \,
H_{\tilde \cube} \, e^{s \left(\sum_{j} H_{\cube_j} - t H_{\cube_0}
\right) } \right\| \nn \\
&\le& \| H_{\tilde \cube} \| \,,
\end{eqnarray}
which is estimated as in \eqref{ma8}.
\medskip

If one works under condition {\bf H3} the proof is similar, using
estimate \eqref{ma9}.  This concludes the proof of Theorem
\ref{likepressure}. \qed

\begin{theorem} \label{momgen2} 
If any of the conditions ${\bf H1}$, ${\bf H2}$, or ${\bf H3}$ hold,
and $\omega$ is a Gibbs-KMS state for $\Phi$, then
\begin{equation}
\lim_\ltoi \left| \frac{1}{|\Lambda|} \log \omega \left(e^{\alpha
K_\Lambda} \right) - \frac{1}{|\Lambda|} \log \frac{ \trace (e^{\alpha
K_\Lambda} \, e^{-H_\Lambda} )} { (e^{-H_\Lambda} )} \right| \,=\, 0
\,.
\end{equation}
\end{theorem}

\proof We will give two different proofs of Theorem \ref{momgen2}. The
first uses the Gibbs condition and not, a priori, the fact that we are
in a one-phase region.  

Defining $W_\Lambda$ as in \eqref{W_L}, we apply the Gibbs condition
\eqref{gibbs} for $\omega$ to the observable $e^{\alpha K_\Lambda}$:
\begin{equation} \label{ma10}
\omega^{-W_\Lambda} \left( e^{\alpha K_\Lambda} \right) \,=\,
\omega^{(\Phi)}_\Lambda \left( e^{\alpha K_\Lambda}\right)
\omega'({\bf 1}) \,=\, \frac{\trace\left( e^{\alpha K_\Lambda}
e^{-H_\Lambda}\right)} {\trace\left( e^{-H_\Lambda}\right)}\,.
\end{equation}
On the other hand, Theorem \ref{infmatrix} asserts that
\begin{equation} \label{ma12}
\left| \log \omega^{-W_\Lambda} \left(e^{\alpha K_\Lambda} \right) -
\log \omega \left(e^{\alpha K_\Lambda}\right) \right| \,\le\, \|
W_\Lambda \| + \sup_{0\le t \le 1} \, \sup_{-\frac12 \le s \le
\frac12} \left\| \tau^{-tW_\Lambda}_{is} (W_\Lambda) \right\| \,.
\end{equation}
By proposition \ref{anal1}, then, if $\Phi$ is in the 
the high temperature regime and $|s| < 1/2$,
\begin{equation}
\frac{1}{|\Lambda|} \left\| \tau^{-tW_\Lambda}_{is}(W_\Lambda)
\right\| \,\le\, \frac{1}{1 - \frac{\lambda}{4} \|\Phi\|_\lambda}
\frac{1}{|\Lambda|} \sum_{X \cap \Lambda \not= \emptyset \atop X \cap
\Lambda^{c} \not= \emptyset } \| \phi_{X} \| \, e^{\lambda |X|} \,,
\end{equation}
which vanishes when $\ltoi$, as we have checked in \eqref{ma11}. The
same, of course, happens to $|\Lambda|^{-1} \| W_\Lambda \|$. Putting
together \eqref{ma10}, \eqref{ma12} and the last two estimates
proves the theorem in the case {\bf H1}.

If $\bf{H2}$ applies, we have only two relations that have to do with
the specific case at hand: the rest of the proof are algebraic
manipulations for KMS states. The first relation is
\begin{equation} \label{ma13}
\tau_s \,=\, \tau_s^{(\Phi' + \Phi'')}  \,=\, \tau_s^{(\Phi')} \circ\,
\tau_s^{(\Phi'')} 
\end{equation}
(the notation should be clear), and the second is
\begin{equation} \label{ma14}
\tau_s^{(\Phi'')} (W_\Lambda) = W_\Lambda \,.
\end{equation}
Eq.~\eqref{ma13} comes from the fact that $\left[H'_V \,,\, H''_V
\right] = 0$, for all finite sets $V \subset \bZ^d$, and that $\tau_s$
is the limit of finite-volume dynamics. As concerns Eq.~\eqref{ma14},
we define
\begin{equation} \label{W_LV}
W_\Lambda(V) \,=\, \sum_{ X \subset V \atop X \cap \Lambda \not=
\emptyset \,;\, X \cap \Lambda^{c} \not= \emptyset} \phi_{X} \,.
\end{equation}
As in the proof of Theorem \ref{likepressure}, $\left[ H''_V \,,\,
W_\Lambda(V) \right] = 0$, so, taking again the limit $V \nearrow
\bZ^d$, and noting that $\| W_\Lambda(V) - W_\Lambda \| \to 0$, we
derive \eqref{ma14}.

Now, using \eqref{ma13} and \eqref{ma14} in \eqref{taup}, we get that,
for the perturbed dynamics,
\begin{equation} 
\tau_s^{-tW_\Lambda} (A) \,=\, \tau_s^{(\Phi'),\, -tW_\Lambda}
\left( \tau_s^{(\Phi'')}(A) \right) \,.
\end{equation}
We plug $A = W_\Lambda$ in the above, exploit \eqref{ma14} again, and
take the analytic continuation of the result: for $|s| \le 1/2$,
\begin{equation} 
\tau_{is}^{-tW_\Lambda} (W_\Lambda) \,=\, \tau_{is}^{(\Phi'),\,
-tW_\Lambda} (W_\Lambda)
\end{equation}
which is estimated as in case ${\bf H1}$. 

One proceeds similarly when ${\bf H3}$ holds. This concludes the first
proof of Theorem \ref{momgen2}.
\medskip

The second proof is based on the fact that---as we have thoroughly
recalled earlier---the Gibbs-KMS state is unique, under our
assumptions. Therefore we can write $\omega$ as limit of finite-volume
Gibbs states with free boundary conditions:
\begin{equation}
\omega(A) \,=\, \lim_{V \nearrow \bZ^d} \frac{\trace(A \, e^{- H_V})}
{\trace(e^{-H_V})} \,,
\end{equation}
for $A \in \bigcup_X \calO_X$. Let us write $H_{V} = H_{\Lambda} +
H_{\Lambda^{c}} + W_\Lambda(V)$, where $W_\Lambda(V)$ was defined in
Eq.~\eqref{W_LV}.  If $A\in \calO_{\Lambda}$, with $\Lambda \subset
V$,
\begin{equation}
\frac{\trace ( A e^{- H_V})} {\trace (e^{-H_V})} = \frac{\trace ( A
e^{-H_\Lambda})} {\trace (e^{ - H_\Lambda} )} \, \frac{\trace ( A
e^{-H_\Lambda - H_{\Lambda^c} - W_\Lambda(V)} )} {\trace ( A
e^{-H_\Lambda - H_{\Lambda^c}}) } \, \frac{\trace (e^{ - H_\Lambda -
H_{\Lambda^c}})} {\trace (e^{ - H_\Lambda - H_{\Lambda^c} -
W_\Lambda(V)}) } \,,
\label{dec} 
\end{equation}
because the trace factorizes, when evaluating the product of two
observables with disjoint support.  Now, via Lemma \ref{matrix}, a
couple of estimates of the type seen in Theorem \ref{likepressure}
yield
\begin{eqnarray}
&& \lim_\ltoi \frac{1}{|\Lambda|} \left| \log \frac{\trace (e^{ -
H_\Lambda - H_{\Lambda^c}})} {\trace (e^{ - H_\Lambda - H_{\Lambda^c}
- W_\Lambda(V)})} \right| \,=\, 0 \,, \nn \\
&& \lim_\ltoi \frac{1}{|\Lambda|} \left| \log \frac{ \trace ( A
e^{-H_\Lambda - H_{\Lambda^c} - W_\Lambda(V)})} {\trace ( A
e^{-H_\Lambda - H_{\Lambda^c}}) } \right| \,=\, 0 \,,
\end{eqnarray}
uniformly in $A \in \calO_\Lambda$, $A>0$, and in $V \supset \Lambda$.
Thanks to this uniformity, one obtains the assertion of Theorem
\ref{momgen2} from \eqref{dec}.  \qed

We conclude by proving what we have called our main result.
\medskip

\noindent {\em Proof of Theorem \ref{main}:~} Combining Theorems
\ref{likepressure} and \ref{momgen2} we have that
\begin{eqnarray} 
f^{(\Psi,\Phi)}(\alpha) &=& \lim_\ltoi \frac{1}{|\Lambda|} \log
\omega( e^{\alpha K_\Lambda} ) \nn \\
&=& \lim_\ltoi \frac{1}{|\Lambda|} \log \trace(e^{\alpha K_\Lambda}
e^{-H_\Lambda }) - \lim_\ltoi \frac{1}{|\Lambda|} \log
\trace(e^{-H_\Lambda }) \nonumber \\
&=& g^{(\Psi,\Phi)}(\alpha) -P(\Phi) \,.
\end{eqnarray}
The existence of the pressure is of course a standard result not
harder than Theorem \ref{likepressure}.

The convexity of $f^{(\Psi,\Phi)}$ follows from the convexity of
$\alpha \mapsto \log \omega(e^{\alpha K_\Lambda})$, which is verified
with a standard application of H\"older's inequality, noting that
$\omega(e^{\alpha K_\Lambda}) = \int d\nu(x) \, e^{\alpha x}$, for
some Borel measure $\nu$ (coming from the spectral measure of
$K_\Lambda$ in the GNS representation).

To obtain the Lipschitz continuity, we apply Lemma
\ref{matrix} with $H=\alpha_2 K_\Lambda$, $P= (\alpha_1-\alpha_2)
K_\Lambda$, and $C = e^{-H_\Lambda}$. Since $H$ and $P$ commute,
\begin{eqnarray}
&&\bi \frac{1}{|\Lambda|}\left| \log \trace \left( e^{\alpha_1
K_\Lambda} e^{-H_\Lambda} \right) - \log \trace \left( e^{\alpha_2
K_\Lambda} e^{-H_\Lambda} \right) \right| \nn \\
&\le& \frac{1}{|\Lambda|} \| (\alpha_1 -\alpha_2) K_\Lambda \| \,\le\,
|\alpha_1 -\alpha_2| \sum_{X \ni 0} \|\psi_X\| \,,
\end{eqnarray}
which easily leads to \eqref{lip}. \qed

\begin{acknowledgments} 
We thank Jean-Pierre Eckmann, Henri Epstein, Giovanni Gallavotti,
Vojkan Jaksic, Claude-Alain Pillet, Charles-Edouard Pfister, and
Lawrence E.~Thomas for their comments, encouragement, and
suggestions. This work has its origin in the joint work of one of us
(M.~L.) with Joel Lebowitz and Herbert Spohn, to whom we are
particularly grateful.  L.~R.-B.\ acknowledges the support from NSF
Grant DMS-0306540.  M.~L.\ received travel funds from NSF (Block Grant
DMS-0306887), G.N.F.M.\ (Italy) and COFIN--MIUR (Italy).
\end{acknowledgments}

\end{document}